\documentclass[aps,nofootinbib,preprint]{revtex4}
\usepackage{bm}
\usepackage{amssymb}
\usepackage{lscape}
\usepackage{amsmath}
\usepackage{graphicx}
\usepackage{epsfig}
\begin{document}

\title{Traversable wormhole magnetic monopoles from Dymnikova Metric.}
\author{ $^{2}$ Jes\'us Mart\'{\i}n Romero\footnote{E-mail address: jesusromero@conicet.gov.ar}, $^{1,2}$ Mauricio Bellini\footnote{E-mail: mbellini@mdp.edu.ar}}

\address{$^{1}$ Departamento de F\'{\i}sica, Facultad de Ciencias Exactas y
Naturales, Universidad Nacional de Mar del Plata, Funes 3350,
(7600) Mar del Plata,
Argentina.\\
$^2$ Instituto de Investigaciones F\'{\i}sicas de Mar del Plata (IFIMAR). Consejo Nacional de Investigaciones Cient\'{\i}ficas y T\'ecnicas (CONICET). }

\begin{abstract}
We study a traversable wormhole originated by a transformation over the 4D Dymnikova metric which describes analytic Black-Holes (BH). By using a transformation of coordinates which is adapted from the used in the Einstein-Rosen bridge, we study a specific family of geodesics in which a test particle with non-zero electric charge induces an effective magnetic monopole, that is perceived by observers outside the wormhole. Because the Riemannian geometry cannot explain the presence of magnetic monopoles, then we propose a torsional geometry in order to explore the possibility that magnetic monopoles can be geometrically induced. We obtain an expression that relates torsion and magnetic fields jointly with a Dirac-like expression for magnetic and electric charges, such that torsion makes possible define a fundamental length that provides a magnetic field and a spacetime discretization.
\end{abstract}


\maketitle
\section{Introduction and motivation}\label{secintro}

Up to now, General Relativity (GR) has passed every experimental tests. The more recent was
the prediction of the wave-forms of gravitational waves coming from the merger of two black holes
(BH)\cite{1}, and two Neutron stars\cite{2}, which were detected since 2015 by the LIGO-VIRGO collaboration. Since the first
detection, many other BH-BH mergers have been detected, in agreement with expectations of General
Relativity\cite{3,4,5,6}. The problem, today not resolved is that at very little scales the theory is plagued by singularities\cite{7,8,9}, that indicate the breakdown of its
predictivity at very small distances. It is manifested at the center of black holes. The more known case is the Schwarzschild metric, but it is expected that quantum gravity can solve this important problem.

In order to resolve this problem, there were many attempts. The first author to propose a non singular black hole solution was Bardeen\cite{n7}. Later, some authors
developed other proposals\cite{n8,n9,n11,n12,n13,n14,n15,n19,n20,n23}. Their scheme was generalized by I. Dymnikova\cite{n10} for spherical and static solutions, and in the frame of Nonlinear Electrodynamics Coupled to Gravity\cite{n10b}. Further properties can be
found in\cite{n27}. Wormholes was proposed by Flamm\cite{Flamm} in 1916, after introduction of
General Relativity. H. Weyl\cite{hweyl} also speculated about the wormhole existence in 1928, but the first serious calculations were done by Einstein and Rosen \cite{4einsteinrosen} in 1935, in order to obtain an atomistic theory of matter and electricity. This was named the “Einstein-Rosen Bridge”. The term wormhole was introduced by Wheeler\cite{wheeler} some years later. The Einstein-Rosen bridge was constructed to connect two Schwarzschild black-holes, but the presence of a singularity in the bridge implies that this is not-traversable. Recent progress has shown that traversable wormholes may exist, being real physical objects\cite{real,real1,real2,real3,real4}. Such advances open a very interesting research area which deserves be explored. Traversable wormholes produce violations of the null energy condition, which is only possible quantum mechanically.
Then, one may expect that such wormholes must be of quantum size or must be surrounded by exotic matter. An alternative to that picture is to consider the presence of torsion as a geometrical entity which "can provide how exotic the wormhole is" \cite{torswh1,torswh2,torswh3}. In this work, we develop a wormhole without singularity and we provide the link to a first approximation to quantization of magnetic and electric charges, once a torsional geometry associated to the wormhole is assumed in the sense of Weiztenb\"{o}ck geommetry\cite{6weit}. Furthermore, we study the Dymnikova proposal for a black-hole in order to study the wormhole associated to this metric and the possibility to geometrically induce a magnetic charge using the torsional structure coefficients.

This paper is organized as follows:
In Sect. (\ref{sec1}) we present some basic elements about the Dymnikova spacetime, which describes an analytic black-hole. We include some basic considerations about what is a regular black-hole, and we mention the meaning, units and value of the parameters which are present in Dymnikova metric. In Sect. (\ref{sec2}) we develop a transformation in which is based in the Einstein-Rosen bridge, but adapted to the Dymnikova black-hole. We study the transformed metric and we obtain an induced analytic wormhole with finite throat. Furthermore, we introduce a special family of geodesics that represent orbits along the throat of the wormhole, which are of special interest. In Sect. (\ref{sec3}) we study the magnetic field originated by an electric charge that orbits the wormhole's throat and is observed by an exterior observer. The magnetic field agrees qualitatively with that of a magnetic monopole with an effective magnetic charge, that is related to electric charge by a Dirac-like discretization. In Sect. (\ref{sec4}) we study a torsional geometry which is compatible with the induced magnetic monopole. Because the Riemann geometry cannot explains the presence of magnetic monopoles without a singularity, then we introduce a geometry capable to explain the observed monopole, that, after using a Weitzenb\"{o}ck geometry, we obtain a new connection which is defined in terms of vierbein. These vierbein are related to the trajectory of the test charge, and the  torsion related to the Weitzenb\"{o}ck connections, enable us discretizate magnetic and electric charges. Finally, in Sect. (\ref{sec5}) we present final considerations and remarks.\\

\section{Some elements about Dymnikova Metric}\label{sec1}

We started from the Dymnikova metric\cite{n11}, which describes an analytic black hole with a line element given by
\begin{eqnarray}\label{1.1dymnikovametric}
ds^{2}\,=\,f(r)\,dt^{2}\,-\,f(r)^{-1}\,dr^{2}\,-\,r^{2}\,\left(\,d\theta^{2}\,+\,\sin ^{2}\theta\,d\phi^{2}\,\right),
\end{eqnarray}
with
\begin{eqnarray}\label{1.2f}
f(r)\,=\,1\,-\,\frac{ab}{r}\,\left(\,1\,-\,e^{\frac{-r^{3}}{a^{3}}}\right),
\end{eqnarray}
in which $a$ and $b$ are parameters. The parameter $a$ has dimensions of length and $b$ is dimensionless. This line element is a de Sitter-like spacetime near the center and resembles the Schwarzschild metric with total mass $ m\,=\,\frac{a\,b}{2}$ at large distances. Null radial curves in which $ds^{2}=0$, $d\theta=0$ and $d\phi=0$ defines what is called {\it a coordinate speed of light}\cite{2visser}, according to \begin{eqnarray}\label{1.3c}
c(r)\,=\,\left|\frac{dr}{dt}\right|\,=\,|f(r)|.
\end{eqnarray}
An effective refractive index is defined by $\eta(r)\,=\,\frac{1}{|f(r)|}$. Such index remains finite when $c(r)\,\neq\, 0$. The Dymnikova metric guarantee that for $b\,<\,b_{crit}=1.456$, the function $f(r)$ does not have zeros\cite{3poncedeleon,4poncedeleon}. Therefore, it is easy to see from the equation (\ref{1.3c}), that $c(r)\,\neq\, 0$ represents a regular black hole. We are going to impose condition $b\,=\,1\,<\,b_{crit}$ for economy reasons, without loss of generality.

\section{Dymnikova Wormhole.}\label{sec2}

The usual transformation to obtain Einstein-Rosen bridge\cite{4einsteinrosen},\cite{5MBJRWH} is defined by
\begin{eqnarray}\label{2.1trafo}
u^{2}\,=\,r\,-\,r_s,
\end{eqnarray}
where $u$ is the new "radial" coordinate and $r_s$ is the minimum radius in the new metric. Such transformation mathematically collapses the interior of the BH to a single point at the zero of the new coordinate. When we are dealing with a Schwarzschild BH, then $r_s$ is the Schwarzschild radius, but the Dymnikova metric is well defined in all the domain: $r_s\,=\,0\,\leq\,r\,<\,\infty$. Therefore the transformation given in eq.(\ref{2.1trafo}) is reduced to
\begin{eqnarray}\label{2.2notrafo}
u^{2}=r.
\end{eqnarray}
We can define the throat area for a wormhole by
\begin{eqnarray}\label{2.3area}
A(u)\,=\,4\,\pi\,{r^{2}(u)},
\end{eqnarray}
which, once we use the transformation defined in Eq. (\ref{2.2notrafo}), results in that the minimum throat area is $A(0)=0$, breaking with the effective traversability of the obtained wormhole.\\

We must modify the Eq. (\ref{2.2notrafo}) to be of the shape of Eq. (\ref{2.1trafo}), in order to obtain a finite minimum throat area, for that we need to introduce a length $l_0$:
\begin{eqnarray}\label{2.3sitrafo}
u^{2}\,=\,r\,-\,l_0,
\end{eqnarray}
which guarantees that $A(u)\,=\,4\,\pi\,{\left(u^{2}\,+\,l_0\right)}^{2}$
and then the minimum throat area remains finite
\begin{eqnarray}\label{2.4miniarea}
A_{min}\,=\,A(0)\,=\,4\,\pi\,l_0^{2}.
\end{eqnarray}
The wormhole metric obtained from the Dymnikova metric is described by the line element
\begin{equation}\label{2.5whmetric}
ds^{2}\,=f(u)\,dt^{2} - \frac{1}{f(u)}\,du^{2}\,-\left(u^{2}\,+\,l_0\right)^{2}\,\left(\,d\theta^{2}\,+\,\sin^{2}\theta\,d\phi^{2}\,\right),
\end{equation}
where
\begin{equation}
f(u)=\left[1\,-\,\frac{a}{u^{2}\,+\,l_0}\left(1\,-\,e^{-\,\left(\frac{u^2\,+\,l_0}{a}\right)^{3}}\right)\right],
\end{equation}
such that $u\,=\,\left(-\infty\,,\,  +\infty\right)$. Here, one obtains $u(r)\,=\,0$ for $r\,=\,l_0$. The metric of (\ref{2.5whmetric})
excludes a part of the manifold which is inside the ball of radius $r\,=\,l_0$. Therefore, we need $l_0$ to be a small quantity. In order to propose a value for $l_0$, we must notice that for a big $l_0$, most of the manifold will remain out of our mathematical description, which is not desirable. Therefore, there are physical arguments to chose a small and finite $l_0$, in order to introduce a fundamental length in the theory (maybe Plankian length). We must notice that $l_0$ plays an analogous role to the Schwarzschild radius in the Einstein-Rosen bridge, but there aren't singularities in the Dymnikova metric. Of course, $l_0$ must be of the order of Plank length, when we want describes quantum wormholes. The throat of traversable wormholes has been extensively studied by Visser and Hochberg in\cite{Visser}, where they analyze the energy conditions and qualitative properties. On the another hand, the parameter $a$ is related to the mass of a Schwarzschild BH that could be very significant. Then, we assume $l_0\, \ll \,a $. The geodesics are defined by the equations
\begin{eqnarray}\label{geogen}
\frac{\partial^2\,t}{\partial\,\lambda^2}\,&+&\,\frac{12\,u^6\,e^{-u^6}\,+\,e^{-u^6}\,-\,1}{u\,\left(\sqrt{u}\,-\,1\,
+\,e^{-u^6}\right)}\,\frac{\partial\,t}{\partial\,\lambda}\,\frac{\partial\,u}{\partial\,\lambda}=0, \nonumber \\
\frac{\partial^2\,u}{\partial\,\lambda^2}\,&+&\,-\,\frac{\left(\sqrt{u}\,-\,1\,+\,e^{-u^6}\right)\left(12\,u^6\,e^{-u^6}\,
+\,e^{-u^6}\,-\,1\right)}{16\,u^4}\left(\frac{\partial\,t}{\partial\,\lambda}\right)^2\, \nonumber \\
&+&\,\frac{12\,u^7\,e^{-u^6}\,
+\,4\,u^{\frac{3}{2}}\,+\,5\,u\,e^{-u^6}\,-\,5\,u}{\left(\sqrt{u}\,-\,1\,+\,e^{-u^6}\right)\,u^2}\,\left(\frac{\partial\,u}{\partial\,\lambda}\right)^2\,-\,\frac{1}{2}\,
\left(u\,-\,\sqrt{u}\,+\,\sqrt{u}\,e^{}-u^6\right)\left(\frac{\partial\,\theta}{\partial\,\lambda}\right)^2\, \nonumber \\
&-&\,\frac{1}{2}\,\left(u\,-\,\sqrt{u}\,
+\,\sqrt{u}\,e^{-u^6}\right)\,\sin(\theta)^2\,\left(\frac{\partial\,\phi}{\partial\,\lambda}\right)^2=0, \nonumber \\
\frac{\partial^2\,\phi}{\partial\,\lambda^2}\,&+&\,\frac{4}{u}\,\frac{\partial\,u}{\partial\,\lambda}\,\frac{\partial\,\phi}{\partial\,\lambda}\,
+\,\frac{2\,\cos(\theta)}{\sin(\theta)}\,\frac{\partial\,\theta}{\partial\,\lambda}\,\frac{\partial\,\phi}{\partial\,\lambda}=0, \nonumber \\
\frac{\partial^2\,\theta}{\partial\,\lambda^2}\,&+&\,\frac{4}{u}\,\frac{\partial\,u}{\partial\,\lambda}\,\frac{\partial\,\theta}{\partial\,\lambda}\,
-\,\sin(\theta)\,\cos(\theta)\,\left(\frac{\partial\,\phi}{\partial\,\lambda}\right)^2=0,
\end{eqnarray}
in which  $t\equiv t(\lambda)$, $u\equiv u(\lambda)$, $\theta \equiv \theta(\lambda)$, and $\phi \equiv \phi(\lambda)$. The metric of (\ref{2.5whmetric}) admits a special family of geodesics in which the affine parameter is $\lambda\,=\,t$  and the coordinate $u$ is constant $u\,=\,0$. Under such assumptions the system of differential equations (\ref{geogen}) is very simplified and admits at least some special solutions. Two of the equations become trivial and the other two are very simple. The pair of remaining equations describes the behaviour of $\theta$ and $\phi$. These geodesics are described by curves situated over some kind of casket spire located at the throat of the wormhole according with: \begin{eqnarray}\label{2.6geodesics}
0&=&\frac{\partial^{2}\,\phi}{\partial\,\lambda^{2}}\,+\,2\,\left(\frac{\cos(\,\theta)}{\sin(\,\theta)}\right)\,\frac{\partial\,\theta}{\partial\,\lambda}
\,\frac{\partial\,\phi}{\partial\,\lambda},\\\nonumber 0&=&\frac{\partial^{2}\,\theta}{\partial\,\lambda^{2}}\,-\,\sin(\,\theta)\,\cos(\,\theta)\,\frac{\partial^{2}\,\phi}{\partial\,\lambda^{2}}.
\end{eqnarray}
A particular solution with constant angular velocity is admitted: $\theta,\,\phi\,\sim\,\lambda$, by taking into account that such coordinates are compact and cyclical (we use natural units along present paper).\\

\section{Effective magnetic monopole.}\label{sec3}

Magnetic monopoles are the magnetic charges which originates divergent magnetic fields, and therefore plays an analogous role than the electric charge for the electric field. A magnetic monopole must be the center of a magnetic radial field which is decaying with the square of the distance and proportional to the magnetic charge magnitude. The presence of magnetic monopoles preserves the symmetry of the Maxwell equations by duality transformations between electric and magnetic fields. First modern mention of a magnetic monopole was made in 1931 by P. A. M. Dirac who demonstrated that the mere existence of magnetic charges grants that both electric and magnetic charges are quantized\cite{4.1dirac}. In 1968, Wu-Yang showed that when a central monopole is obtained from a defined potential field, it must obeys some conditions under gauge transformations, which grants charge quantization\cite{4.2wu}. In 1974 G. T'hooft proved that great unification theories imply the presence of massive magnetic monopoles\cite{4.3thooft}. Magnetic monopoles are very interesting elements that were studied and are still investigated. They are present in condensed-matter theories and in string theory. However, our approach is different to those, and has a geometrical nature.

The charged particle which is following geodesic of Eq. (\ref{2.6geodesics}), moves on a trajectory $l$, which is restricted to $u\,=\,0$ and enclose the interior of the ball with radius $r\,=\,l_0$, that is the part of the original manifold that was neglected by metric (\ref{2.5whmetric}), in the throat of the wormhole. Accomplishing with
\begin{eqnarray}\label{3.1dl}
dl\,=\,\underbrace{l_u\,du}_{=0}\,+\,l_\theta\,d\theta\,+\,l_\phi\,d\phi,
\end{eqnarray}
the particle moves with velocity $\mathbf{v}\,=\,v_\theta\,d\theta\,+\,v_\phi\,d\phi$. We can take a look in a fixed moment over the surface given by $u\,=\,0$, $t\,=\,t_0$. We consider metric tensor associated to the line element (\ref{2.5whmetric}). We can specialize on the surface described by $u\,=\,0$, $t\,=\,t_0$, in order to obtain a two-dimensional metric. We shall refer $\mathbf{g}_2$ to the metric tensor that describes this surface, and we shall denote its determinant with $g_2$. In this framework, the magnetic field generated by some unitary magnetic charge is
\begin{eqnarray}\label{3.2campo}
\mathbf{B}^{(u)}&=&|v|\,\oint\,\left[\frac{v_\theta\,d\theta\,+\,v_\phi\,d\phi}{S^{2}(l,u)}\,\wedge\,dl\right]\,\sqrt{g_2}\,=\\\nonumber &=&\oint\,\left[\frac{v_\theta\,l_\phi\,-\,v_\phi\,l_\theta}{S^{2}(l,u)}\right]\,\mathbf{\varepsilon}^{\theta\phi (u)}.
\end{eqnarray}
The $2$-form $d\theta\,\wedge\,d\phi$ can be understood as parallel to $du$ at a fixed time. Then $\mathbf{\varepsilon}^{\theta\phi}(u)\,du\,=\,\sqrt{g_2}\,d\theta\,\wedge\,d\phi|_l\,=\,\sqrt{g_2}\,\left(d\theta\,\wedge\,d\phi|_l\right)^{(u)}du$ will be the length form over the casket surface. Outside the wormhole, when $|u|\,\gg\, l_0$, we obtain that $S^{2}(l,u)$ is given by
\begin{eqnarray}\label{maso}
S^{2}(l,u)\,=\,|u|^{2},
\end{eqnarray}
such that $|u|$ is the distance from the source to the point in which we are measuring the field. Obviously the obtained magnetic field is a $1$-form, and must point out in $u$-direction. Then, the components of the magnetic field due to the magnetic monopole, will be
\begin{eqnarray}\label{3.3campo}
\mathbf{B}^{u}\,=\,\frac{|v|}{|u|^{2}}\,\oint\,\left(v_\theta\,l_\phi\,-\,v_\phi\,l_\theta\right)\mathbf{\varepsilon}^{u\theta\phi}.
\end{eqnarray}
The term between parenthesis in  (\ref{3.3campo}) is the angular momentum of the field source particle with respect to the point in which is earlier located the particle. The spacetime discretization\cite{9desa} is necessary as an approach to spacetime quantization. It could be made by linking torsion, which provides a geometric description, and using the Plank length. To discretize the spacetime, we must descretize the amount between parenthesis in (\ref{3.3campo}). Supported by previous considerations we could assume the spacetime discretization. Therefore, the effective magnetic charge can be quantized [we shall return over spacetime discretization in Sect. \ref{sec4}]:
\begin{eqnarray}\label{3.4campo}
|\mathbf{B}|\,\sim\,\frac{n}{|u|^{2}}\,=\,\frac{q_m}{|u|^{2}},
\end{eqnarray}
where $\mathbf{n}^{(u)}\,=\,\oint\,\left(v_\theta\,l_\phi\,-\,v_\phi\,l_\theta\right)\mathbf{\varepsilon}^{\theta\phi (u)}$, $\mathbf{n}\,=\,n\,\check{u}$. For a source given by an electric charge $q_e$, we obtain that
\begin{eqnarray}\label{3.5quant}
q_m\,=\,n\,q_e,
\end{eqnarray}
which describes a Dirac-like discretization between a true electrical charge $q_e$ and the effective magnetic charge $q_m$, which can be observed from outside the wormhole.
Therefore, outside of the wormhole, an $u$-dial magnetic field which is decaying with $\frac{1}{|u|^{2}}$ can be detected. Of course this field is not decaying as $\frac{1}{r^{2}}$, and then can be related to the magnetic monopole only in the transformed spacetime outside the wormhole. The Riemann geometry does not support magnetic monopoles when the system of coordinates is coordinated. However, we have showed that a magnetic monopole can be induced as an effective entity by a transformation which excludes a part of the original manifold, creating a topological defect. Such exclusion looks to be not physical but only a mathematical trick, because there is no singularity to hide. Then, there is not physical reasons to interpret $l_0$ as a true event horizon. Therefore, the magnetic monopole will be accessible because is located on the throat of a traversable wormhole. However, in the proximity of the throat [see Eq. (\ref{maso})], an observer notes that we are dealing with an electric current spire. Notice that, in the present example, no magnetic monopoles are located outside the wormhole. \\

\section{On torsional geometry and quantization.}\label{sec4}

In this section, we shall provide an interpretation of the geometrical nature of the effective magnetic monopoles. As we shall see, they can be linked to a torsional geometry, which is the simplest one that can be given by a construction, and is similar to those of Weiztenb\"{o}ck geommetry\cite{6weit}.The Weiztenb\"{o}ck geometry is characterized by a connection which is defined by the vierbein, such that its covariant derivative is null and the torsion can be nonzero. The Weitzenb\"{o}ck geometry is a paradigmatic example of a geometry which is characterized by a connection defined in terms of the vierbein, which relate two distinct basis. In most of cases, one of the basis is chosen to be normalized but it can be non-coordinate, and therefore, the Weitzenb\"{o}ck geometry will describe torsion, which is really a consequence of the structure that appears under a transformation of coordinates. This fact motivates us to build a torsional geometry over the casket $u\,=\,0$. We are going to interpret the (\ref{3.4campo}) as an integral of a specific torsion, in which the elements $l_\theta,\,l_\phi,\,v_\theta,\,v_\phi$ can be associated to a vierbein and their derivatives.

\subsection{Vierbein, $2$-dimensional part of the transformation.}\label{subsec4.1}

The $2$-dimensional part of the transformation and the vierbein interpretation can be obtained from the way in which the test particle is moving over the geodesics restricted to the casket. We must start by defining a convenient basis: $\{\underrightarrow{E}_\alpha\}\,=\,\{d\theta,\,d\phi\}$, and a basis of the cotangent space (of the casket with $u\,=\,0$ in any given point). We can define a new basis of the cotangent space at the same point: $\{\underrightarrow{e}^i\}\,=\,\{e^3,\,e^4\}$, according to \begin{eqnarray}\label{4.1coord}
e^3&=&\underbrace{l_\theta}_{e^3_\theta}\,d\theta\,+\,\underbrace{l_\phi}_{e^3_\phi}\,d\phi,\\\label{4.2coord}
e^4&=&e^4_\theta\,d\theta\,+\,e^4_\phi\,d\phi,
\end{eqnarray}
in which we selected $e^3\,=\,dl$ as the direction pointing along curve $l$, and $e^4$ is determined up a factor, by $\mathbf{g}_2\left(e^3,\,e^4\right)\,=\,\left[\mathbf{g}_2\right]_{\theta\theta}\,e^3_\theta\,e^4_\theta\,
+\,\left[\mathbf{g}_2\right]_{\phi\phi}\,e^3_\phi\,e^4_\phi\,=\,0$. The Eqs.(\ref{4.1coord}) and (\ref{4.2coord}), can be expressed as
\begin{eqnarray}\label{4.3vier}
\underrightarrow{e}^i\,=\,e^i_\alpha\,\underrightarrow{E}^\alpha,
\end{eqnarray}
where objects $e^i_\alpha$ are called vierbein\cite{7Einsteinyepez}.\footnote{Latin indexes $i$ and $j$ are taking values $i,\,j\,=\,3,\,4$ (chosen notation is going to be clear soon) and Greek indexes $\alpha$ and $\beta$ are running over values $\alpha,\,\beta\,=\,\theta,\,\phi$.}
The analog relation to Eq. (\ref{4.3vier}) for the corresponding dual basis of the tangent space, is given by \begin{eqnarray}\label{4.4vier}
\overrightarrow{e}_{i}\,=\,e_i^\alpha\,\overrightarrow{E}_{\alpha},
\end{eqnarray}
where we must remember that duality is expressed by $\underrightarrow{e}^i\left(\overrightarrow{e}_j\right)\,=\,\delta^i_j$, and $\underrightarrow{E}^\alpha\left(\overrightarrow{E}_\beta\right)\,=\,\delta^\alpha_\beta$. Therefore, we can say that vierbein relating basis of casket's tangent space are determined by $e_i^\alpha\,e_\alpha^j\,=\,\delta^j_i$, $e_i^\alpha\,e_\beta^i\,=\,\delta^\beta_\alpha$.  The speed of the source particle is defined as a cotangent vector by \begin{eqnarray}\label{4.1speed}
\mathbf{v}\,=\,\underbrace{\frac{dl_\theta}{dt}}_{v_{\theta}=\frac{\partial e^3_\theta}{\partial t}}\,d\theta\,+\,\underbrace{\frac{dl_\phi}{dt}}_{v_{\phi}=\frac{\partial e^3_\phi}{\partial t}}\,d\phi.
\end{eqnarray}
Of course, $\mathbf{v}$ is a vector which is co-tangent to the $1$-dimensional manifold given by curve $l$.

\subsection{Vierbein, $4$-dimensional extension.}

It is easy to extend previous construction to both, $4$-dimensional tangent and cotangent spaces, which consider the casket as part (trivially embedded) of the $4$-dimensional spacetime. Therefore, we obtain \begin{eqnarray}\label{4.2W1}
\{\widetilde{\underrightarrow{e}}\}&=&\{dt,\,du,\,e^3,\,e^4\},\\\label{4.2W2}\{\widetilde{\underrightarrow{E}}\}&=&\{dt,\,du,\,d\theta,\,d\phi\}.
\end{eqnarray}
We must consider a few more vierbein: $e^1_t\,=\,e^2_u\,=\,1$. The set of vierbein can be written as $\{e^n_\nu\}$ and $\{e^\mu_m\}$. The bases and vierbein are related by
\begin{eqnarray}\label{4dbasevier}
\widetilde{\underrightarrow{e}}^n&=&e^n_\nu\,\widetilde{\underrightarrow{E}}^\nu,\\\nonumber\widetilde{\underrightarrow{e}}_n
&=&e_n^\nu\,\widetilde{\underrightarrow{E}}_\nu,\\\nonumber \delta^n_m&=&e^n_\mu\,e^\mu_m,\\\nonumber \delta^\mu_\nu&=&e^m_\nu\,e^\mu_m.
\end{eqnarray}

\subsection{Connections.}\label{subsec4.3}

The Weitzenb\"{o}ck connections can be expressed as
\begin{eqnarray}\label{4.3wcone}
^{(W)}\Gamma^{\rho}_{\mu\nu}\,=\,e^{\rho}_{n}\,\frac{\partial e^{n}_{\mu}}{\partial x^\nu},
\end{eqnarray}
that are related with a Weitzenb\"{o}ck torsion
\begin{eqnarray}\label{4.3tor}
^{(W)}T^{\rho}_{\nu\mu}\,=\,e^{\rho}_{n}\,\frac{\partial e^{n}_{\mu}}{\partial x^\nu}\,-\,e^{\rho}_{n}\,\frac{\partial e^{n}_{\nu}}{\partial x^\mu}.
\end{eqnarray}
The field of (\ref{3.3campo}) can't be explained in terms of such torsion, but keeps a great resemblance. This is the reason by which we choose a vierbein based connection definition. In order to provide to the equation (\ref{3.3campo}) a geometric interpretation, is necessary to propose a more simple (but still torsional) geometry, in which connections are still given by the vierbein, but according to
\begin{equation}\label{4.4cone}
\Gamma^{a}_{bc}\,=\,{T^{a}_{bc}}\,+\,K^{a}_{bc}.
\end{equation}
Here, $K^{a}_{bc}\,=\,\frac{\partial e^{3}_{a}}{\partial t}e^{3}_b\,h^{a}$ are the contortion tensor components, with $h^{a}|_{a=u}=1$ and $\left.h^{a}\right|_{a=t,\theta,\phi}=0$. The relevant components are:
\begin{eqnarray}\label{4.5cone}
K^{u}_{\phi\theta}&=&\frac{\partial e^{3}_{\phi}}{\partial t} e^{3}_{\theta},\nonumber \\
K^{u}_{\theta\phi}&=& \frac{\partial e^{3}_{\theta}}{\partial t} e^{3}_{\phi}.
\end{eqnarray}
Notice that the connections that differs from the Riemannian ones, are $\Gamma^{u}_{\theta\phi}=\{^{u}_{\theta\phi}\}+K^{u}_{\theta\phi}$, and $\Gamma^{u}_{\phi\theta}={T^{u}_{\phi\theta}}+K^{u}_{\phi\theta}$. All other connections match with the Riemannian ones. Therefore, the geometry is torsional, with
\begin{eqnarray}\label{4.6tor}
T^{u}_{\phi\theta}\,=\,\frac{\partial e^{3}_{\theta}}{\partial t}e^{3}_\phi\,-\,\frac{\partial e^{3}_{\phi}}{\partial t}e^{3}_\theta\,=\,-T^{u}_{\theta\phi},
\end{eqnarray}
which are the only non-vanishing torsion terms. The torsion is present, but geodesics in the equation (\ref{2.6geodesics}) remains unchanged. From the equation (\ref{3.3campo}), after some algebraic manipulation, we obtain
\begin{eqnarray}\label{4.7term}
\mathbf{B}^{u}&=&\frac{|v|}{|u|^{2}}\,\oint\,\left(v_\theta\,l_\phi\,-\,v_\phi\,l_\theta\right)\,\mathbf{\varepsilon}^{u\theta\phi}\,=\\\nonumber &=&\frac{|v|}{|u|^{2}}\,\oint\,T^{u}_{\phi\theta}\,\mathbf{\varepsilon}^{\theta\phi},
\end{eqnarray}
which is the only nonzero component. The last integral could be understood as the Burgers vector\cite{8burgers}. In material science, when mapping an ideal crystal into a crystal containing a fracture, Burgers vector represents the magnitude and direction of a lattice dislocation. Burgers vector can be interpreted as originated in torsional geometry and defined by:
\begin{eqnarray}\label{4.8burg}
\oint\,T^{u}_{\phi\theta}\,\mathbf{\varepsilon}^{\theta\phi}\,=\,b^u.
\end{eqnarray}
V. de Sabbata and B. Kumar Datta\cite{9desa} have studied the physical meaning of such defect term, by including the interpretation of torsion as a magnetic spin, which is of course quantized. Therefore, will be reasonable to suppose that the components of the Burgers vector could be an integer number of Planck lengths, that is $b_m\,=\,n_m\,l_p$, in which $n_m$ is an integer numbers (we use natural units). The former expression can be adapted to the context of the present paper, in which the magnetic field is entity related to the torsion.
Therefore, we obtain $|b|\,=\,|b_u|\,=\,n\,q_e$, in which $n$ is zero or a positive integer. This argument is supporting discretization that we have used in Sect. (\ref{sec3}), in order to obtain the Eq. (\ref{3.4campo}) and the quantization in Eq. (\ref{3.5quant}).

\section{Final Remarks}\label{sec5}

We have introduced and studied the case of a traversable wormhole obtained from Dymnikova spacetime, in which there is a charged test particle in a specific kind of geodesics located in the throat of the wormhole. The Riemannian geometry is not sufficient to explain the presence of monopoles in a spacetime which is free of singularities. Therefore, by using a Weintzenb\"{o}ck geometric representation, we have obtained that it induces an effective interior magnetic monopole for any observer outside the wormhole, so that both charges meet the Dirac's quantization condition.

In Sect. (\ref{subsec4.3}) we have mentioned that the Burgers vectors can be associated with Plank length, in order to give a geometrical nature to the constant. This is not convenient in the present work, in which the obtained torsion is related to the magnetic field. However, it can be fixed by the choice $l_0\,=\,n_l\,l_p$, that could be an optional manner to introduce the Planck length to the theory as a magnitude originated by the geometry. In this manner the Planck length would be being introduced from a transformation described in equation (\ref{2.3sitrafo}), and by restricting the wormhole to the quantum domain.

We have studied an effective magnetic monopole which is obtained as the manifestation of an electric charge in the throat of the traversable wormhole. An inverse picture is obtained from the same scenario, with the only difference that we supposed a magnetic monopole test particle, and therefore from the outside of the wormhole is viewed as an electric charge. The last assertion can be easily obtained from Sect. (\ref{sec2}), by changing electric and magnetic charge. This is consistent with the results obtained by us in a previous work in which we described a wormhole linking two effective 4D induced black-holes. We induced it from 5D spacetime by doing a specific set of foliations obtaining that a not accessible magnetic(electric) charge cited in the interior of the wormhole can be intended as an effective electric(magnetic) charge outside the induced 4D black-hole\cite{5MBJRWH}.

The Riemmanian geometry (that we have employed in this work), does not support the magnetic monopoles existence without a topological defect, due to the fact $d(d(A))\,=\,0$.
However, in our present case we have obtained an effective magnetic monopole with one unreal singularity, because the Dymnikova metric is analytic. That can be explained because the used transformation removes a part of the original manifold in a such manner that it is restricted to the throat of the wormhole, that includes an effective torsional defect at the origin, for the observer located outside the wormhole.

\section*{Acknowledgements}

\noindent The authors acknowledge CONICET, Argentina (PIP 11220150100072CO) and UNMdP (EXA852/18) for financial support.

\end{document}